\documentclass[3p]{elsarticle}

\usepackage{amssymb}

\usepackage{textcomp}
\usepackage{gensymb} 
\usepackage{subfig}  

\usepackage{makecell} 
\usepackage{hyperref} 

\usepackage{graphicx}
\usepackage{balance}
\usepackage{color}
\usepackage{xcolor}
\usepackage{multirow}
\usepackage{caption}
\usepackage{epsfig,amsfonts,amsmath,cases,amssymb,graphics, latexsym, times, algorithmic, algorithm, bbm, color, soul,wrapfig,comment}

\journal{Smart Health}

\begin{document}

\begin{frontmatter}



\title{Visualization of Emergency Department Clinical Data for Interpretable Patient Phenotyping\tnoteref{tnote1}}
\tnotetext[tnote1]{This paper is under consideration at Smart Health as part of the IEEE/ACM Conference on Connected Health: Applications, Systems and Engineering Technologies. \\
\textcopyright 2019. This manuscript version is made available under the CC-BY-NC-ND 4.0 license http://creativecommons.org/licenses/by-nc-nd/4.0/
}


%

\author[tamu]{Nathan C. Hurley\corref{cor1}}
\ead{natech@tamu.edu}
\cortext[cor1]{Corresponding author: 
3112 TAMU, HRBB, College Station, TX, USA
  Tel.: +1-979-458-3870;}

\author[yale]{Adrian D. Haimovich}

\author[yale]{R. Andrew Taylor}

\author[tamu]{Bobak J. Mortazavi}

\address[tamu]{Department of Computer Science and Engineering, Texas A\&M University, United States}

\address[yale]{Department of Emergency Medicine, Yale School of Medicine, United States}


\begin{abstract}

Visual summarization of clinical data collected on patients contained within the electronic health record (EHR) may enable precise and rapid triage at the time of patient presentation to an emergency department (ED).  The triage process is critical in the appropriate allocation of resources and in anticipating eventual patient disposition, typically admission to the hospital or discharge home. EHR data are high-dimensional and complex, but offer the opportunity to discover and characterize underlying data-driven patient phenotypes. Data-driven phenotypes are intended to relieve reliance on weak labels like diagnosis codes and to aid in identifying populations of existing patients that are most similar to a specific patient.  These phenotypes will enable improved, personalized therapeutic decision making and prognostication. In this work, we focus on the challenge of two-dimensional patient projections.  A low dimensional embedding offers visual interpretability lost in higher dimensions. While linear dimensionality reduction techniques such as principal component analysis are often used towards this aim, they are insufficient to describe the variance of patient  data. This linear reduction does not account for higher order, non-linear interactions of variables. In this work,  we employ the newly-described non-linear embedding technique called uniform  manifold  approximation  and  projection (UMAP).  UMAP seeks to capture both local and global structures in high-dimensional data. We then use Gaussian mixture models to identify clusters in the embedded data and use the adjusted Rand index (ARI) to establish stability in the discovery of these clusters.  This technique is applied to five common clinical chief complaints from a real-world ED EHR dataset, describing the emergent properties of discovered clusters. We observe clinically-relevant cluster attributes, suggesting that visual embeddings of EHR data using non-linear dimensionality reduction is a promising approach to reveal data-driven patient phenotypes.  In the five chief complaints, we find between 2 and 6 clusters, with the peak mean pairwise ARI between subsequent training iterations to range from 0.35 to 0.74.

\end{abstract}


\end{frontmatter}

\section{Introduction}
  \label{sec:intro}

Electronic health records (EHRs) include heterogeneous data that represent past and ongoing patient care episodes. The EHR is accessed as both a real-time information transfer environment as well as a medium for retrospective analysis. In the emergency department (ED) setting, patients are, in the vast majority of cases, first seen by medical professionals for registration and triage. This process links a digital record to the individual waiting to be seen and enables the rapid assessment of patient complexity as well as the visit urgency.

ED triage is made more challenging by increases in patient volume \cite{ed_trends} and relative subjectivity in the triage process \cite{triage_scales}. The emergency severity index (ESI) is a five-level triage system developed to improve robustness in nurse-driven triage assessments and has been shown to correlate with admission rate and mortality \cite{esi}. However, this approach is not designed to leverage the breadth of data available in the EHR. More recently, various machine learning approaches to augmenting triage have been described \cite{data}\cite{deep_triage}\cite{ml_triage} and there is mounting evidence to suggest that incorporating heterogenous data into initial patient assessments enables a more refined and accurate prediction of critical patient outcomes. 

While risk and event prediction is now a mature field within medical informatics \cite{cv_ml}, there is growing interest in leveraging massive EHR databases to uncover phenotypes of complex disease processes. These efforts have multiple aims which include the automated discovery of patient populations for retrospective analyses and the identification of specific subgroups that may benefit from particular interventions or therapies \cite{ahmad2018}. Prior work has shown potential utility in varied fields including hematology \cite{comp_clin_pheno} and cardiology \cite{ahmad2018}.

In the ED setting, however, there is a pressing need for tools that enable prospective and interpretable phenotyping. Visualization offers one appealing approach to interpretability and techniques like t-distributed stochastic neighbor embedding (t-SNE) \cite{tsne} have been used in varied settings with great success \cite{visne}. However, a given visualization produced by t-SNE is unable to be expanded to future data points, as t-SNE produces a non-parametric visualization \cite{tsne}.  In contrast, uniform manifold approximation and projection (UMAP) is a parametric visualization technique that preserves more global structure than does t-SNE while also allowing future data points to be fit to an existing model without recreating the entire model \cite{umap}.

Here, we describe the first application of EHR phenotype visualization to the ED triage process. Using a database containing 560,486 anonymized patient visits with 972 sparsely-populated features, we have previously shown the ability to robustly predict patient disposition (hospital admission or discharge) using a very small subset (n = 15) of these features\cite{data}. In that work, we found that an XGBoost model trained on triage score, medication counts, demographics, and hospital usage statistics could predict hospital admission with an AUC of 0.91.  That work found that although models trained on triage data or history data performed similarly, models trained on both triage and history data showed a marked improvement in predicting admission.  However, that work focused on the binary outcome of admission or discharge.
In the present work, we expand on this prior work by implementing and validating a technique to visualize subpopulations within this dataset.  We show that new patients can be mapped into this visualization, and we show that patient similarities and differences to local subpopulations can give information about likely clinical outcomes in order to aid clinical assessment.  We anticipate that this work could be utilized in a clinical setting in order to aid in understanding relationships between patients and to aid in clinical decision making.

The contributions of this work are:
\begin{itemize}
    \item A method of using UMAP for non-linear dimensionality reduction, Gaussian mixture models (GMMs) for clustering data, and a combination for data visualization.
    \item A metric utilizing the adjusted Rand index (ARI) between different folds of data clustering to determine appropriate model hyperparameters and cluster stability in random subsets of the data.
    \item An application in real-world clinical data of patients presenting with five common clinical chief complaints.  The emergent properties of these clusters are described, and clinically-relevant attributes are discussed to show clinical decision support potential.
\end{itemize}
The rest of this paper is organized as follows.  In section \ref{sec:related_work}, we discuss related work in patient phenotyping and dimensionality reduction.  We also discuss the utilization of ARI as a metric for measuring partition similarity.  In section \ref{sec:methods}, we discuss the data used and the process of building our model to find and validate clusters within the clinical dataset.  In section \ref{sec:results} we show the model embeddings and results on both the synthetic and clinical dataset.  We discuss the clinical characteristics of the clusters discovered within the clinical dataset.  In section \ref{sec:discussion}, we discuss the significance of the results and some of the clinical pictures that can be drawn from the results.  We then discuss directions for future applications of this work.

\section{Related Work}
  \label{sec:related_work}

Previous patient phenotyping work has focused on identifying phenotypes among patients with a given disease state, such as heart disease \cite{chf_cluster_anal}, sepsis \cite{sepsis_pheno}, or amyotrophic lateral sclerosis \cite{bk_ehr_pheno}.  In these works, authors have used various clustering techniques, such as hierarchical agglomeration after reducing dimensionality with principal component analysis (PCA) \cite{chf_cluster_anal} or by using K-means clustering on patient clinical severity scores \cite{sepsis_pheno}.  Clusters have also been discovered through training a semi-supervised denoising autoencoder, and then using PCA or t-SNE to represent the hidden nodes of the autoencoder and identifying the clusters manually \cite{bk_ehr_pheno}.  In contrast, our approach is not specifically tailored to a given disease state, but rather is applied to any patient presenting for emergency care with a particular chief complaint.

Other work has looked at enumerating a wide range of specific phenotypes, and then identifying those phenotypes within a patient population \cite{phekb}.  However, these approaches are supervised techniques; physicians with top-level domain knowledge drive the phenotype discovery.  In these approaches, the phenotype is identified clinically, and new patients are matched to the phenotype clusters using custom rules and logic. In contrast, our approach is not trained with a particular clinical condition or outcome in mind, but instead searches for phenotype clusters within a given patient set.  This allows for application of our technique to new and unseen phenotypes without the need for expert evaluation.

Dimensionality reduction techniques are often used in visualizing multidimensional data.  Earlier approaches have used PCA and GMMs to visualize populations of samples in a 2D space in order to aid in clinical diagnosis \cite{pca_to_gmm}.  PCA has also been used in conjunction with other clustering methods such as agglomerative clustering or K-means clustering \cite{pca_gene_data}.  However, in sparse datasets PCA often reaches a limit where it is unable to express data without losing a significant amount of information about the dataset variability.

Other sparse clinical datasets have been visualized with t-SNE \cite{visne}. t-SNE is a powerful tool that embeds data into lower dimensions while maintaining structure present at higher dimensions \cite{tsne}.  However, t-SNE is non-parametric; while it is effective at embedding a given dataset, it is difficult to embed new, previously unseen members of that dataset. UMAP is a newer dimensionality reduction technique that is able to scale beyond t-SNE while also expressing relations that t-SNE is unable to express \cite{umap}.  UMAP is a parametric approach, and as such is able to embed new data without necessitating a retraining of the model.  Although several studies have utilized UMAP for phenotyping cell populations \cite{bm_umap}\cite{s_cell_umap}, we did not find any studies that utilize UMAP for patient phenotyping with EHR data.

The adjusted Rand index (ARI) is a widely-used metric of similarity between two partitions of a given set of objects \cite{adjusted_rand}.  The ARI for two partitions of a set ranges from a value of 1, when the two partitions are equivalent, to near 0 when the two partitions are chosen at random.  The closer the ARI of two partitions is to 1, then the more similarly partitioned the set is.  ARI is robust to partitions of different sizes, and has been used for evaluating the results of various clustering techniques \cite{ari_clustering_pca}.

\section{Methods}
  \label{sec:methods}

\begin{figure}[!t]
    \centering
    \includegraphics[width=0.9\textwidth]{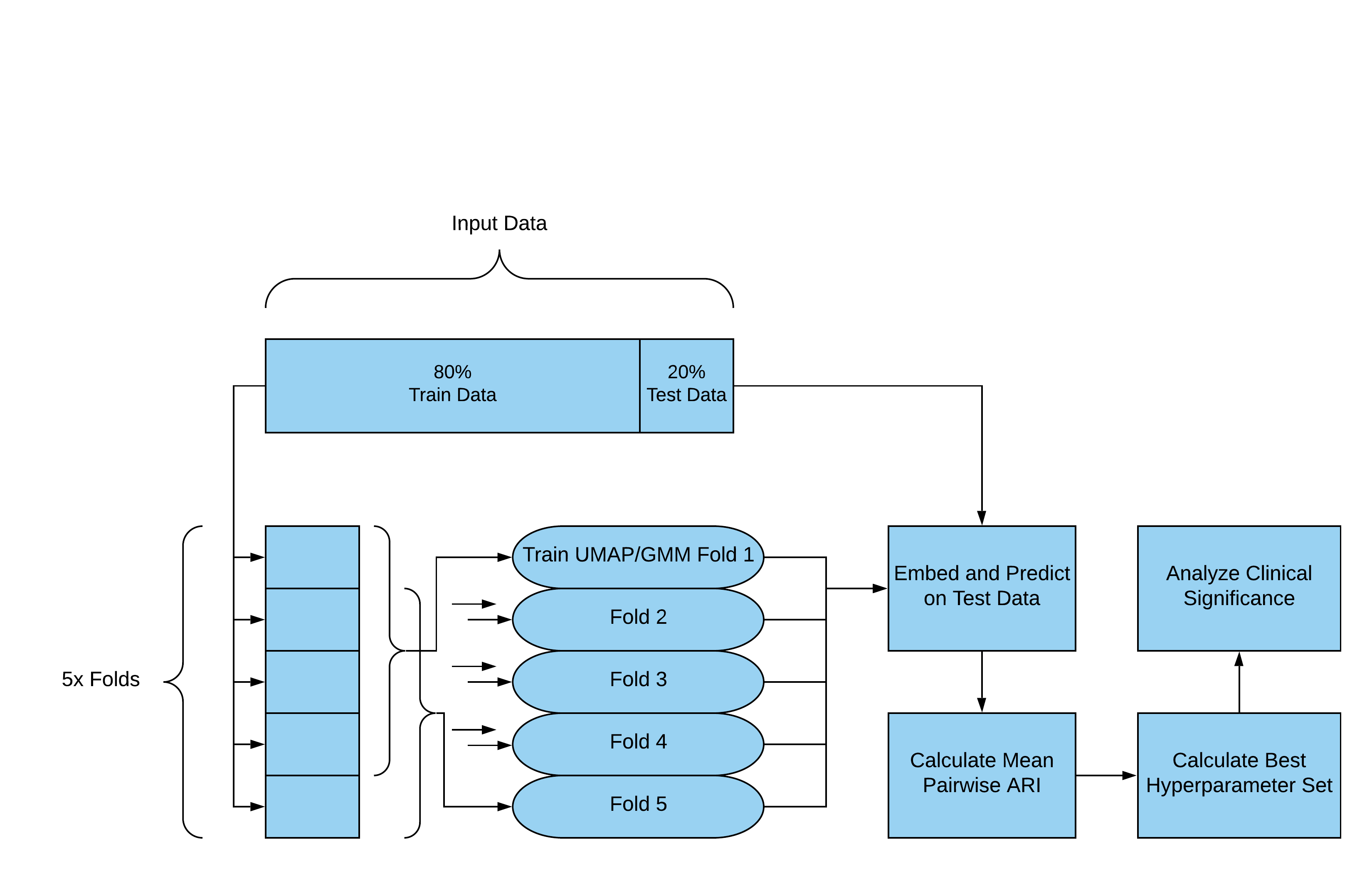}
    \caption{A diagram of the method presented here.  The data is randomly split into training data (80\%) and testing data (20\%).  The training data is split into five folds.  Each combination of four folds is used to train a separate UMAP -> GMM model, which is then applied to the testing data.  The mean pairwise ARI is calculated between each test data cluster prediction, and the set of hyperparameters giving the best agreement is selected for clinical analysis.}
    \label{fig:method}
\end{figure}

In this section, we describe our visualization technique.  The objective is to visualize an embedding of EHR data which preserves global structure so that physicians can rapidly infer relationships between new patients and well-defined subpopulations of patients.  We first filter our data by chief complaint upon presentation to the ED.  Next, we split our data into a training set and a validation set.  The training set is split into five partitions which are embedded in two dimensions using UMAP.  GMMs are trained on each partition, and then the model trained on each partition is applied to the validation set.  The number of clusters in the final model is determined by maximizing the ARI among labels for the validation set.  A diagram of this process in shown in Figure \ref{fig:method}. We perform this process both on synthetic data to demonstrate the method's viability, and on a clinical dataset to demonstrate the clinical applicability.  All code used here is available online at \url{https://github.com/nch08a/EDVizPhenotyping}.

\subsection{Datasets}
\subsubsection{Synthetic Data}
A synthetic dataset was generated using the make\_classification package in scikit-learn \cite{scikit-learn}.  The dataset was generated with 100,000 samples of 100 features, 50 of which were informative.  The dataset was generated to have 10 classes, and for the class separation to be 0.75.  These parameters were chosen to generate a dataset with noise and a large amount of high-dimensional structure.  Having 50 informative features prevents the data from being easily explained by any simple projection to a 2D space.  Utilizing non-informative features adds noise to the model in a way similar to correlated patient data.  A class separation of 0.75 results in overlapping distributions, so that samples do not clearly align with one cluster or another.  With this overlap, there may be some samples which are not able to be correctly classified even with full model information.  This noise allows the dataset to reflect difficulties in clinical datasets.

\subsubsection{Clinical Data}
The clinical dataset used in this study was previously detailed and is publicly available \cite{data}. This dataset includes 560,486 patient visits at three EDs, collected from March 2014-July 2017.  This dataset was preprocessed to include all adult patients who were either discharged or admitted.  Data collected about the patients includes disposition, triage evaluation, chief complaint, hospital usage, past medical history, outpatient medications, historical labs and vitals, and imaging/ekg counts.  The dataset used in this paper is available online as described in \cite{data}.

Early analysis did not show cluster stability when the method was applied to the entire dataset, and so the data was broken down into subsets by patient chief complaint. The five largest chief complaint subsets were analyzed here, but this technique could be applied to any chief complaint subset.  The five subsets analyzed here were the only subsets consisting of at least 3\% of the total dataset. Statistics relating to the patient chief complaint subpopulations are shown in Table \ref{table:pop}.  

\begin{table}
\caption{Selected Patient Characteristics}
\scriptsize     
\centering
\begin{tabular}{lcccccccccc}
\hline
                                                            & \multicolumn{2}{c}{Abdominal Pain} & \multicolumn{2}{c}{Chest Pain} & \multicolumn{2}{c}{Shortness of Breath} & \multicolumn{2}{c}{Back Pain} & \multicolumn{2}{c}{Falls} \\
                                                            \hline
Count                                                       & 54,315         &                    & 35,778        &                 & 24,652             &                     & 20,633       &                 & 19,012     &               \\
Disposition- Admit (\%)                                     & 19,482         & 35.9             & 16,065        & 44.9\%          & 15,791             & 64.1\%              & 3,061        & 14.8\%          & 5,642      & 29.7\%     \\
Gender--- Male (\%)                                         & 19,169         & 35.3\%             & 16,587        & 46.4\%          & 10,231             & 41.5\%              & 9,010        & 43.7\%          & 7,823      & 41.2\%        \\
Insurance Status--- Medicare (\%)                           & 20,833         & 38.4\%             & 11,778        & 32.9\%          & 6,530              & 26.5\%              & 8,690        & 42.1\%          & 4,677      & 24.6\%        \\
Insurance Status--- Medicaid (\%)                           & 8,957          & 16.5\%             & 8,376         & 23.4\%          & 9,426              & 38.2\%              & 3,234        & 15.7\%          & 6,890      & 36.2\%        \\
Language--- English (\%)                                    & 48,560         & 89.4\%             & 32,479        & 90.8\%          & 22,861             & 92.7\%              & 18,537       & 89.8\%          & 17,811     & 93.7\%        \\
Arrival via Ambulance (\%)                                  & 12,532         & 23.1\%             & 13,603        & 38.0\%          & 11,470             & 46.5\%              & 4,168        & 20.2\%          & 10,688     & 56.2\%        \\
Mean Age in Years (range)                                   & 45.7          & 18-105             & 53.4         & 18-105          & 61.2              & 18-107              & 47.3        & 18-103          & 61.1      & 18-107        \\
Mean Triage Heart Rate (range)                              & 85.8          & 35-187             & 84.6         & 30-240          & 88.8              & 30-199              & 84.3        & 44-205          & 83.6      & 30-180        \\
Mean Triage Systolic BP(range)                              & 132.5         & 59-248             & 135.7        & 59-274          & 134.7             & 60-312              & 134.0       & 63-246          & 135.0     & 59-261        \\
Mean Triage Diastolic BP (range)                            & 80.6          & 27-194             & 81.8         & 28-172          & 80.2              & 30-214              & 81.3        & 32-157          & 80.2      & 28-156        \\
Mean Triage Respiratory Rate (range)                        & 17.6          & 8-61               & 17.7         & 8-69            & 18.6              & 10-66               & 17.6        & 8-64            & 17.5      & 18-57         \\
Mean Triage O$_2$ Saturation (range)                        & 97.4          & 67-99              & 97.3         & 60-99           & 96.6              & 60-99               & 97.4        & 71-99           & 97.1      & 73-99         \\
Mean Triage Temperature in \degree F (range) & 98.1          & 94.5-104.7         & 98.0         & 94.1-104        & 98.1              & 90.1-106            & 98.0        & 94.3-104.4      & 98.0      & 93.3-103.4    \\
Mean Prior Admissions (range)                               & 1.2           & 0-49               & 1.4          & 0-48            & 1.7               & 0-40                & 0.6         & 0-46            & 0.9       & 0-42          \\
\hline
\end{tabular}
\label{table:pop}
\end{table} 

These patients represent a wide variety of adult patients who presented at an ED.  ED care is difficult as the patient population seen is highly variable: there are acute, emergent cases, as well as patients who come to the ED for more routine care that could be more appropriate in other health care settings.  Therefore, the patients shown in Table \ref{table:pop} represent a wide spectrum from healthy to critically ill.  One proxy for severity is the means by which a patient arrives at the ED; a patient arriving via ambulance is more likely to be sick than a patient who walked or drove themselves.  Number of prior admissions can be a proxy for long term health of a patient.  Patients with chronic illness are more likely to have been hospitalized a larger number of times, while patients without chronic illness are less likely to have been hospitalized a larger number of times.  

\subsection{Data Preprocessing}
Each chief complaint dataset was split into 80\% training and 20\% testing.  The training set was further split into 5 training folds by 5-fold cross validation.  However, the testing folds produced by this cross validation were not used for testing, as the classification metric used (ARI) requires the same test fold for every model trained.  All splits were random with no prior weighting of the datasets. Admission data for the given ED visit and emergency severity index (ESI) were omitted so as to censor the model from the eventual clinical outcome and from direct triage assessment. Within each fold, the categorical data was one-hot encoded.  The numerical training data was normalized to have unit range and zero mean.  Missing data was mean imputed.  The test data was normalized by the same scaling factor that produced the training normalization, and missing data was imputed to the mean of the training data.

\subsection{Dimensionality Reduction and Clustering}
Both datasets were embedded into two dimensions for ease of visualization.  Two methods were examined for dimensionality reduction: PCA and UMAP.  These methods were chosen for their ability to provide a parametric dimensionality reduction technique.  This way, the transformations can be trained and subsequently applied to previously unseen data, allowing for clinical phenotyping of new patients.

UMAP was trained with 2, 15, or 150 neighbors included in the local manifold approximation.  These parameters were chosen to represent a spectrum of distance at which structure is considered- a smaller number of neighbors in the approximation emphasizes local structure, while a larger number of neighbors emphasizes global structure.
The minimum embedding distance between points was set to 0, 0.1, or 0.25.  These parameters were chosen to allow for samples to ``clump" to varying amounts.  Smaller distance between points allows more similar points to be stacked together at the expense of losing their relationship to more distant points.
All distances were calculated using the Euclidean metric.  After training each UMAP model on a given training fold, that model was applied to the unseen test fold.

Full covariance GMMs were trained on the training data using an expectation-maximization algorithm.  The GMMs were trained with number of clusters $n$ ranging from 2 to 20.  These models were then used to predict labels of the testing dataset.  The testing data was clustered once for each training fold.

\subsection{Clustering Analysis}
The different clusterings of testing data were analyzed for stability using the ARI.  The mean pairwise ARI was computed between each test set labeling produced with a given value of $n$.  For some values of $n$, the model failed to find $n$ non-null clusters. The final cluster number was selected by choosing the value of $n$ that maximized mean pairwise ARI while still finding at least a mean of $n-0.5$ clusters.  A high ARI indicates that the clustering process is stable among a given dataset, as certain entries are consistently assigned to the same cluster.  A low ARI indicates that the clustering process is dominated by noise, and that there is a large amount of variability between different folds.

\subsection{Clinical Cluster Analysis}
Clusters may be compared to each other or to a dataset as a whole.  In this analysis, comparisons were performed by calculating the difference of the mean of all normalized variables in a cluster with the mean of all normalized variables in the entire chief complaint test set.  Training data was not used in analysis.  Variables with differences furthest from zero represent the features in the cluster that are most distinct from the rest of the dataset.

\begin{figure*}[!t]
    \centering
    \subfloat[]{\label{fig:synth_pca_unlab} \includegraphics[width=.33\textwidth]{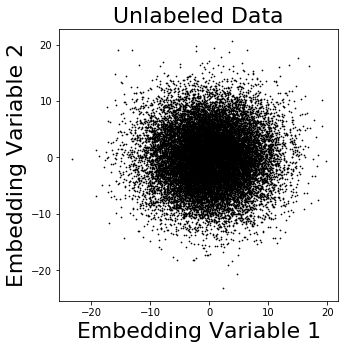}}
    \subfloat[]{\label{fig:synth_pca_ground} \includegraphics[width=.33\textwidth]{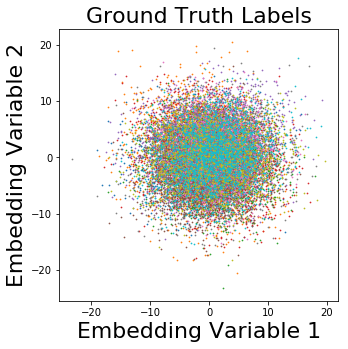}}
    \subfloat[]{\label{fig:synth_pca_predicted}\includegraphics[width=.33\textwidth]{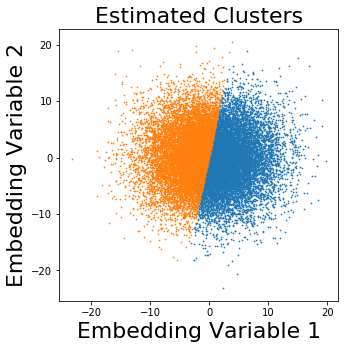}}
    \caption{PCA Embeddings of synthetic data.  Here the embedding and GMMs have been trained on one of the five training splits, and then applied to the test set.  This test set application is shown here.  In \ref{fig:synth_pca_unlab}, the embedded data is shown without labels.  In \ref{fig:synth_pca_ground}, all data has been labeled with the ground truth cluster identities.  In \ref{fig:synth_pca_predicted}, the GMM-predicted clusters are shown.}
    \label{fig:synth_pca}
\end{figure*}

\begin{figure*}[!t]
    \centering
    \subfloat[]{\label{fig:synth_umap_unlab} \includegraphics[width=.33\textwidth]{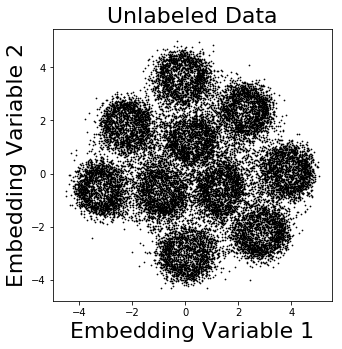}}
    \subfloat[]{\label{fig:synth_umap_ground} \includegraphics[width=.33\textwidth]{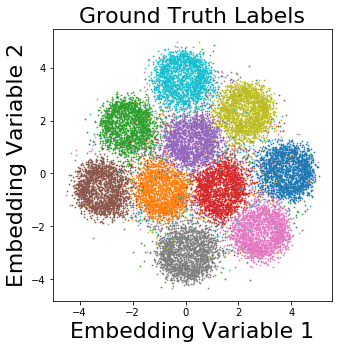}}
    \subfloat[]{\label{fig:synth_umap_predicted}\includegraphics[width=.33\textwidth]{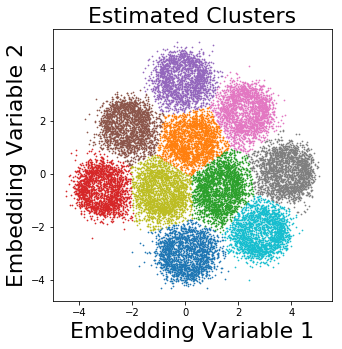}}
    \caption{UMAP Embeddings of synthetic data.  Here the embedding and GMMs have been trained on one of the five training splits, and then applied to the test set.  This test set application is shown here.  In \ref{fig:synth_umap_unlab}, the embedded data is shown without labels.  In \ref{fig:synth_umap_ground}, all data has been labeled with the ground truth cluster identities.  In \ref{fig:synth_umap_predicted}, the GMM-predicted clusters are shown.}
    \label{fig:synth_umap}
\end{figure*}

\begin{figure*}[!t]
    \centering
    \includegraphics[width=\textwidth]{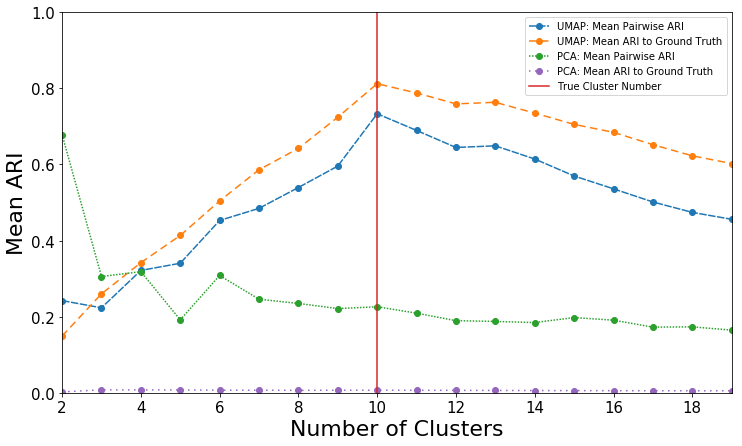}
    \caption{Mean pairwise ARIs of clusterings on synthetic data.  The solid line denotes the true number of clusters.  ARIs are shown both pairwise between different training folds and with respect to the ground truth cluster labeling.}
    \label{fig:synth_summary}
\end{figure*}

\section{Results}
  \label{sec:results}

\subsection{Synthetic Data}
Embeddings of the synthetic data are shown in Figures \ref{fig:synth_pca} and \ref{fig:synth_umap}.  Figure \ref{fig:synth_pca} shows PCA embeddings.  The embeddings used to generate this image were trained on a randomly chosen training split, and applied to the test set.  Figure \ref{fig:synth_pca_unlab} shows the embedding without any cluster labels.  Figure \ref{fig:synth_pca_ground} shows the points labeled with their ground truth cluster identity, and Figure \ref{fig:synth_pca_predicted} shows the most stable prediction from GMMs.  As can be seen by comparing these figures, the GMM here does not show any strong relationship with the true cluster identities.  ARIs of GMMs trained on this data can be seen in Figure \ref{fig:synth_summary}.  Although small cluster number shows high pairwise ARI, no choice of cluster number ever provides a high ARI with respect to the ground truth cluster labels.  These data suggest that PCA struggles to reveal true clusters.

We then trained UMAP embeddings using random training splits on synthetic data (Figure \ref{fig:synth_umap}) and applied the mapping to held-out test data (Figure \ref{fig:synth_umap_ground}).  We then used GMMs to discover the most stable cluster predictions (Figure \ref{fig:synth_umap_predicted}). In contrast to the PCA data in Figure 2, we observed significantly improved capture of phenotypes within the synthetic dataset.  While the embedding does not perfectly separate the clusters, a clear trend towards separation can be seen.  ARIs of the GMMs trained here can be seen in Figure \ref{fig:synth_summary}.  The peak ARI is reached when grouping with 10 clusters, which is the true number of clusters present.

\subsection{Clinical Data}
Within our emergency department electronic health record database, the most common chief complaints were abdominal pain (present in 54,315 patient visits, 9.7\% of total), chest pain (35,778, 6.4\%), shortness of breath (24,652, 4.4\%), back pain (20,633, 3.7\%) and fall (19,012, 3.4\%).  All other chief complaints were present in less than 3\% of patient visits.  Of note, patient visits could have multiple chief complaints, with the average patient visit having 1.13 chief complaints. We sought to implement our embedding and clustering pipeline to visits within each of these categories. We then compared clusters to one another to determine the features driving the phenotype.

\begin{figure*}[!t]
    \centering
    \includegraphics[width=\textwidth]{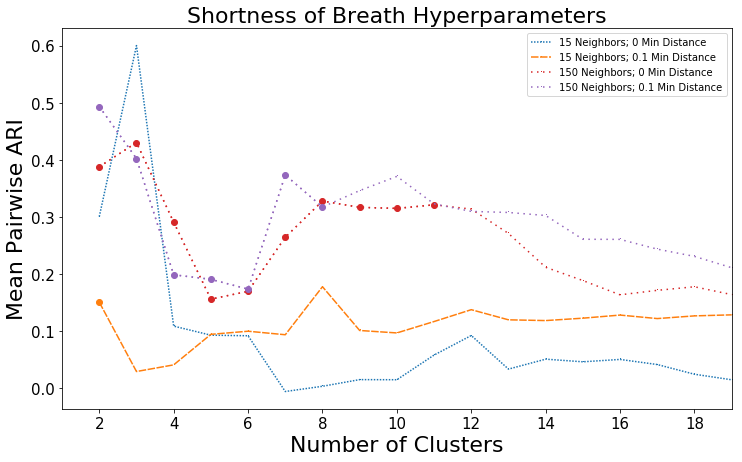}
    \caption{Representative plot of hyperparameters.  Here, four sets of hyperparameters and the resulting mean pairwise ARI are shown.  All ARIs are plotted.  The markers indicate hyperparameters where the mean number of clusters produced was no more than 0.5 less than the number of clusters with which the GMM was trained.  For instance, the blue peak at 3 clusters was built with a model where although 3 clusters were indicated, the mean number of clusters used was 1.2, indicating that four folds categorized all test data as belonging to a single cluster, while one fold categorized all test data as belonging to two clusters.  Therefore, the ARI is elevated through a trivial clustering of only one cluster present.}
    \label{fig:sob_ari}
\end{figure*}

\subsubsection{Shortness of Breath}

We first explored hyperparameters for the Shortness of Breath population (Figure \ref{fig:sob_ari}). In this plot, the highest ARI (15 neighbors, 0 min distance, 3 clusters) is invalidated as a best possible option, as the mean fold produced 1.2 clusters, which is below the cutoff of no more than 0.5 below the number of clusters used for training the model.  The next highest ARI (150 neighbors, 0.1 min distance, 2 clusters) was chosen for further analysis. This analysis was replicated across chief complaints and summarized in Table \ref{table:aris}.

\begin{figure*}[!t]
    \centering
    \subfloat[]{\label{fig:sob_train} \includegraphics[width=.375\textwidth]{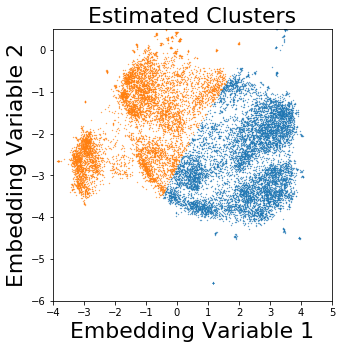}}
    \subfloat[]{\label{fig:sob_test} \includegraphics[width=.375\textwidth]{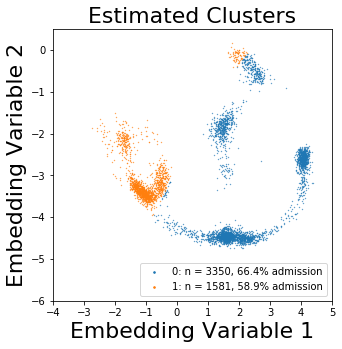}}
    \caption{UMAP embeddings of patients with Shortness of Breath.  Figure \ref{fig:sob_train} shows the training data, while Figure \ref{fig:sob_test} shows the application of the model to the test data.}
    \label{fig:sob}
\end{figure*}

The best clustering of shortness of breath was found with two clusters that contain 72.1\% (95\% CI 67.5-76.7) and 27.9\% (95\% CI 23.3-32.5) of patients.  The larger cluster was slightly more likely to be admitted (66.5\%, 95\% CI 66.3-66.6) while the smaller cluster was less likely to be admitted (57.3\%, 95\% CI 55.3-59.2).  A representative embedding of these clusters are shown in Figure \ref{fig:sob}.

The larger cluster here was much more likely to have been previously visited this hospital system.  These patients were more likely to have arrived via ambulance, suggesting higher acuity.  These patients were slightly more likely to have urinalysis results positive for blood and leukocytes, and were more likely to have risk factors such as chronic obstructive pulmonary disease or congestive heart failure.  On average, these patients had been admitted in this system 2.4 times each previously, with a median of 1 previous admission.

The smaller cluster here was much more likely to have been a first time patient to this ED system.  These patients were more likely to have commercial insurance, and to be employed full time.  These patients were also more likely to present with additional chief complaints, such as cough or palpitations.

\begin{table}[!t]
\caption{Best mean pairwise ARIs and associated hyperparameters per chief complaint.}
\centering
\begin{tabular}{lcccc}
\hline
                    & Best ARI & \# of Neighbors & Minimum Distance & \# of Clusters \\
\hline
Abdominal Pain      & 0.353    & 150          & 0.0       & 2           \\
Chest Pain          & 0.589    & 15           & 0.0       & 6           \\
Shortness of Breath & 0.493    & 150          & 0.1       & 2           \\
Back Pain           & 0.385    & 150          & 0.25      & 4           \\
Falls               & 0.741    & 150          & 0.0       & 2 \\  

\hline
\end{tabular}
\label{table:aris}
\end{table}

\subsubsection{Abdominal Pain}
\begin{figure*}[!t]
    \centering
    \subfloat[]{\label{fig:abd_train} \includegraphics[width=.375\textwidth]{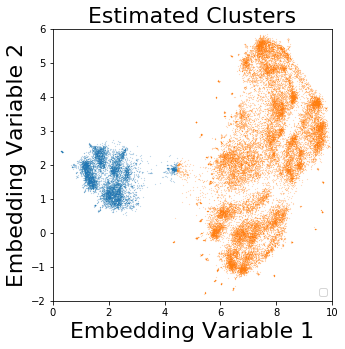}}
    \subfloat[]{\label{fig:abd_test} \includegraphics[width=.375\textwidth]{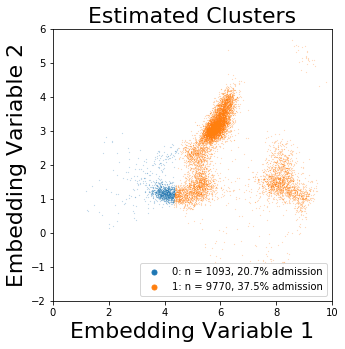}}
    \caption{UMAP embeddings of patients with Abdominal Pain.  Figure \ref{fig:abd_train} shows the training data, while Figure \ref{fig:abd_test} shows the application of the model to the test data.}
    \label{fig:abd}
\end{figure*}

The best clustering of abdominal pain was found with two clusters that contain 89.2\% (95\% CI 83.7-94.6) and 10.8\% (95\% CI 5.4-16.3) of patients (Figure \ref{fig:abd}).  The larger cluster was generally more likely to be admitted (37.6\%, 95\% CI 36.7-38.5), while the smaller cluster was generally less likely to be admitted (24.2\%, 95\% CI 19.3-29.0).  Selected patient characteristics of these clusters are shown and compared in Table \ref{tab:abd}.  The peak mean pairwise ARI found was 0.35, with 2 clusters, 150 neighbors, and no minimum distance between points.

In the smaller cluster, patients tended to have either been previously evaluated in this emergency department system and discharged, or had never been previously evaluated in this system.  These patients had lower blood pressure and less hypertension than the other cluster.  These patients were less likely to have very low $O_2$ saturation, and were less likely to be English speakers.  Relatively few of these patients arrived at the hospital via ambulance, indicating lower acuity.  These patients were less likely to be admitted to the hospital.

The larger cluster tended to have an older population, and more had esophageal disease or hyperlipidemia.  Nearly a quarter of these patients arrived at the hospital via an ambulance, indicating higher acuity.  Most of these patients had been admitted from this ED system before, and over a third were admitted in this visit.

\begin{table}[!t]
\caption{Selected Patient Characteristics of Abdominal Pain Clusters}
\centering
\begin{tabular}{lcccc}
\hline
                                                            & \multicolumn{2}{c}{Cluster 0} & \multicolumn{2}{c}{Cluster 1} \\
\hline
Count                                                       & 1093               & 10.1\%   & 9770        & 89.9\%          \\
Disposition- Admit (\%)                                     & 226                & 20.7\%   & 3667        & 37.5\%         \\
Gender--- Male (\%)                                         & 355                & 32.5\%   & 3488        & 35.7\%          \\
Insurance Status--- Medicare (\%)                           & 539                & 49.3\%   & 3656        & 37.4\%          \\
Insurance Status--- Medicaid (\%)                           & 61                 & 5.6\%    & 1753        & 17.9\%          \\
Language--- English (\%)                                    & 719                & 65.8\%   & 8981        & 91.9\%          \\
Arrival via Ambulance (\%)                                  & 103                & 9.4\%    & 2454        & 25.1\%          \\
Mean Age in Years (range)                                   & 37.8               & 18-89    & 46.8        & 18-103          \\
Mean Triage Heart Rate (range)                              & 83.7               & 43-148   & 86.0        & 40-185          \\
Mean Triage Systolic BP(range)                              & 129.3              & 82-232   & 132.5       & 66-243          \\
Mean Triage Diastolic BP (range)                            & 79.6               & 40-128   & 80.6        & 28-163          \\
Mean Triage Respiratory Rate (range)                        & 17.5               & 13-28    & 17.6        & 10-40           \\
Mean Triage O$_2$ Saturation (range)                        & 97.7               & 91-99    & 97.5        & 74-99           \\
Mean Triage Temperature in \degree F (range) & 98.1               & 96-104.4 & 98.1        & 94.6-103.5      \\
Mean Prior Admissions (range)                               & \textless 0.01  & 0-3      & 1.3         & 0-47            \\
\hline
\end{tabular}
\label{tab:abd}
\end{table}

\subsubsection{Chest Pain}
\begin{figure*}[!t]
    \centering
    \subfloat[]{\label{fig:cp_train} \includegraphics[width=.375\textwidth]{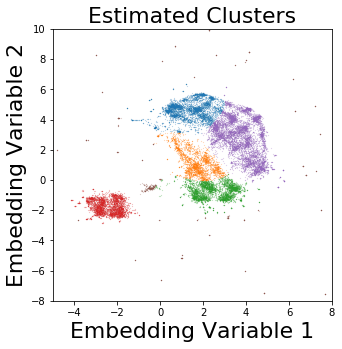}}
    \subfloat[]{\label{fig:cp_test} \includegraphics[width=.375\textwidth]{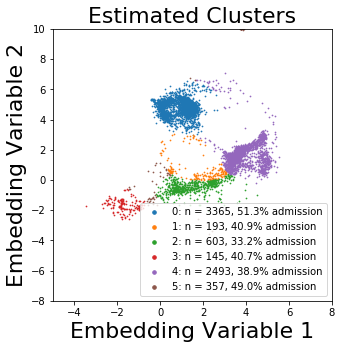}}
    \caption{UMAP embeddings of patients with Chest Pain.  Figure \ref{fig:cp_train} shows the training data, while Figure \ref{fig:cp_test} shows the application of the model to the test data.}
    \label{fig:cp}
\end{figure*}

The best clustering of chest pain was found with six clusters that contain 54.5\% (95\% CI 46.4-62.6), 24.3\% (95\% CI 16.8-31.7), 12.5\% (95\% CI 7.8-17.2), 5.8\% (95\% CI 3.8-7.8), and 0.5\% (95\% CI 0.0-1.0) of patients.  Most clusters had similar rates of admission, with overlapping 95\% confidence intervals.  The peak mean pairwise ARI found was 0.59, with 6 clusters, 15 neighbors, and no minimum distance between points.  A representative visualization of these clusters is shown in Figure \ref{fig:cp}.

In the largest cluster (47.0\% of the population), the patients were slightly more likely to have been previously admitted.  This population was more likely to have hypertension or mood disorders, or to have arrived via ambulance, indicating higher clinical acuity.  These patients were slightly more likely to have positive urine protein and leukocytes. These patients had previously been admitted to this hospital an average of 2.8 times each, with all patients in the top quartile having been admitted 3 or more times each.  51.4\% of patients in this cluster were admitted, as opposed to 44.8\% overall for patients with this chief complaint.

The second largest cluster (34.8\% of the population) was notable for feature patients who were more likely to have never been seen within this hospital system, and very few had ever been admitted.  These patients were more likely white or Caucasian, and were less likely to be on Medicaid (28\%) or Medicare (14\%).  These patients had less hypertension and diagnosed mood disorders than the rest of the population.  38.9\% of these patients were admitted, as opposed to 44.8\% overall.  Patients in this cluster were an average of 6 years younger than the patients in the larger cluster.

The third largest cluster here (8.4\% of the population) was clustered largely on arrival mechanism, and were more likely to have arrived via car or as a walk-in, and were significantly less likely to have arrived via ambulance, indicating a generally lower acuity.  These patients were less likely to have risk factors including male gender, hyperlipedemia, or diagnosed CAD.  These patients were relatively unlikely to be admitted, with only a 33.17\% admission rate.

One cluster contained 5.0\% of the population, and this cluster was more likely to include diabetic patients who had previously been seen in this system.  Patients in this cluster were slightly more likely to have alcohol-related disorders or other substance-related disorders.  These patients also had slightly higher rates of mood or anxiety disorders, and were more likely to have been previously discharged.  Patients in this cluster were admitted at a rate of 49.0\%.

The remaining two clusters each contained less than 3\% of the total population.  Their primary differences relative to the rest of the population were racial demographics.

\subsubsection{Back Pain}
\begin{figure*}[!t]
    \centering
    \subfloat[]{\label{fig:back_train} \includegraphics[width=.375\textwidth]{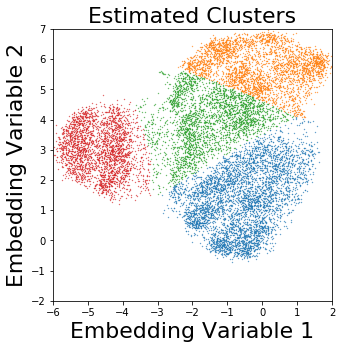}}
    \subfloat[]{\label{fig:back_test}  \includegraphics[width=.375\textwidth]{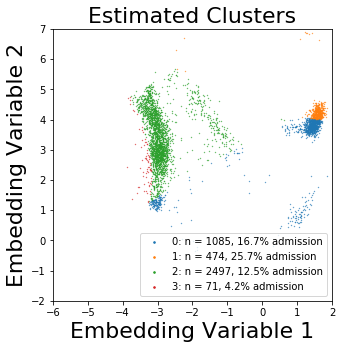}}
    \caption{UMAP embeddings of patients with Back Pain.  Figure \ref{fig:back_train} shows the training data, while Figure \ref{fig:back_test} shows the application of the model to the test data.}
    \label{fig:back}
\end{figure*}

The best clustering of back pain was found with four clusters that contain 63.3\% (95\% CI 60.1-66.6), 22.5\% (95\% CI 20.0-24.9), 12.6\% (95\% CI 10.9-14.3), and 1.6\% (95\% CI 1.3-1.9) of patients.  Clusters had similar rates of admission, with overlapping 95\% confidence intervals.  The peak mean pairwise ARI found was 0.39, with 4 clusters, 150 neighbors, and 0.25 minimum distance between points.  A representative visualization of these clusters are shown in Figure \ref{fig:back}.

The largest cluster, consisting of 60.5\% of the testing data, primarily features patients that were not previously seen in this ED system.  These patients were more likely to be employed full time, and predominantly did not arrive at the hospital via ambulance.  These patients were slightly more likely to have male gender, and were less likely to have asthma.  12.5\% of these patients were admitted, as opposed to 15.0\% of patients with this chief complaint.

The second largest cluster, consisting of 26.3\% of the testing data, features patients that were more often previously seen in this ED system.  These patients were more often insured via Medicaid, and were more likely to have risk factors such as hypertension.  These patients were more likely to have urinalysis positive for leukocytes, blood, or protein.  These patients were more likely to have female gender.  16.7\% of these patients were admitted, as opposed to 15.0\% of the patients overall with this chief complaint.

The next cluster consists of 11.5\% of the testing data.  Patients in this cluster were much more likely to have been previously admitted, and were more likely to have arrived via ambulance.  These patients were generally more hypertensive than the rest of this population, and were more likely to have been diagnosed with mood or anxiety disorders.  These patients were insured with Medicare more often than the remainder of the patients in this population.

The smallest cluster consists of 1.7\% of the testing data.  These patients were predominantly clustered by their racial demographics, and were generally more likely to have arrived at the ED as a walk-in patient.  These patients were slightly more likely to have female gender and to co-present with a chief complaint of having suffered a fall.  Only 4.2\% of these patients were admitted, as opposed to 15.0\% overall.

\subsubsection{Falls}
\begin{figure*}[!t]
    \centering
    \begin{tabular}{cc}
    \subfloat[]{\label{fig:fall_train_0}  \includegraphics[width=.375\textwidth]{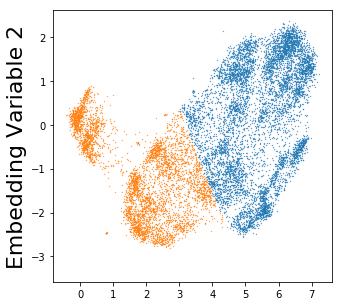}} &
    \subfloat[] {\label{fig:fall_test_0}  \includegraphics[width=.35\textwidth]{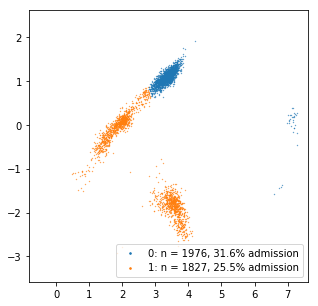}} \\
    \subfloat[]{\label{fig:fall_train_2}  \includegraphics[width=.375\textwidth]{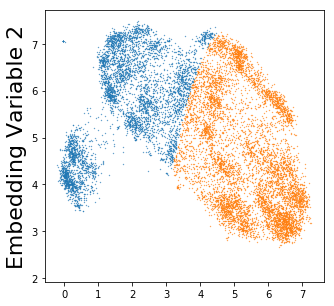}} &
    \subfloat[] {\label{fig:fall_test_2}  \includegraphics[width=.35\textwidth]{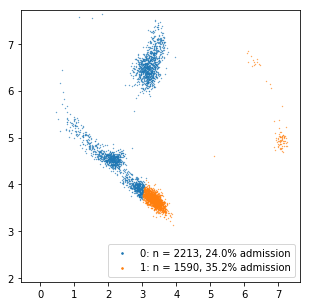}} \\
    \subfloat[]{\label{fig:fall_train_4}  \includegraphics[width=.375\textwidth]{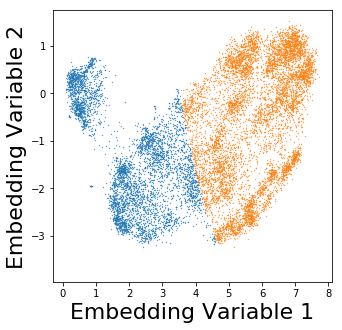}} &
    \subfloat[] {\label{fig:fall_test_4}  \includegraphics[width=.35\textwidth]{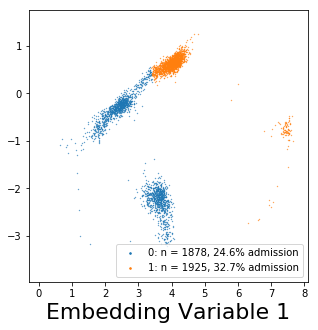}} \\
    \end{tabular}
    \caption{Three different folds of UMAP embeddings of patients who suffered Falls.  Figures \ref{fig:fall_train_0}, \ref{fig:fall_train_2}, and \ref{fig:fall_train_4} show the training data, while Figures \ref{fig:fall_test_0}, \ref{fig:fall_test_2}, and \ref{fig:fall_test_4} show the application of the model to the test data.  In each of these, it can be seen that the embedding follows a similar overall pattern even among different folds.  Two additional folds are not shown, but exhibit the same global shape and cluster characteristics.}
    \label{fig:fall_all}
\end{figure*}

The best clustering of falls was found with two clusters that contain 52.7\% (95\% CI 50.6-54.8) and 47.3\% (95\% CI 45.2- 49.4) of patients.  Clusters had very similar rates of admission. The peak mean pairwise ARI found was 0.74, with 2 clusters, 150 neighbors, and no minimum distance between points.  Three of five representative visualizations of these clusters are shown in Figure \ref{fig:fall_all}.

The larger cluster consists of 52.0\% of the testing data.  In this cluster, patients were more likely to have been previously seen in this ED system.  These patients were more likely to have arrived via ambulance, and were more likely to have positive urinalysis findings for protein, leukocytes, or blood.  These patients were more likely to be on Medicare.  These patients were slightly more likely to have lower blood oxygen saturation.  31.6\% of these patients were admitted, as opposed to 28.7\% overall.

The smaller cluster consists of 48.0\% of the testing data.  In this cluster, patients were more likely to have never been seen in this ED system.  These patients were more likely to be employed full or part time, and to have commercial insurance.  These patients were admitted at a rate of 25.5\%.

\section{Discussion}
  \label{sec:discussion}
In this work, we sought to further efforts towards the summary visualization of high-dimensionality EHR data. We focused on a previously published emergency department dataset with the goal of revealing data-driven phenotypes at time of hospital presentation. We first benchmarked our approach using a synthetic dataset and subsequently moved our analysis towards investigating the five most common emergency department chief complaints.

The synthetic dataset here was generated to have similar size and variability to the clinical dataset.  As was anticipated, two-dimensional PCA was unable to capture the variability contained within the multiple informative dimensions.  We observed very little useable structure within the PCA embedding of the synthetic data (Figure \ref{fig:synth_pca_ground}). While the ARI is higher than would be expected by chance for 2 clusters, it is unremarkable at the true value of 10 clusters.  Furthermore, while the pairwise ARI for the PCA model maintains a value around 0.2, the ARI to the ground truth cluster labels remains at nearly 0.  This suggests that even though some patterns are found within the PCA embedding of the data, these patterns do not correlate with the underlying structure of the data.

In contrast, UMAP clearly identifies elements of the high-dimensional structure of this same dataset.  Even without clustering the UMAP-embedded data, a strong relationship can be seen between most points of a given cluster (Figure \ref{fig:synth_umap_ground}).  Though visually trivial, mapping the mean pairwise ARI of GMMs trained with different numbers of clusters in Figure \ref{fig:synth_summary} shows that the peak ARI correlates with this correct number of clusters. Furthermore, the pairwise ARI and ARI to ground truth are both high values.  This suggests that when using UMAP, the pairwise ARI is useful as a proxy for the ground truth ARI.  When adapting this model to datasets without known ground truth, the pairwise ARI should function to ensure that valid cluster patterns are being observed.
These data motivate our efforts to investigate a real-world clinical dataset. 

\subsection{Clinical Interpretation}
Here, we consider more broadly the clinical implications of the discovered phenotype clusters. In the shortness of breath clusters described above, the characteristics of the two clusters paint two very different clinical images.  In the larger cluster, the characteristics present a description of patients who are generally sick: patients who are high utilizers of healthcare with medical comorbitities such as congestive heart failure or chronic obstructive pulmonary disease.  On the other hand, the smaller cluster represents first-time utilizers of the health care system who are generally healthier.  They are more likely to present with other chief complaints, such as cough.  This suggests that this cluster represents more acute causes of shortness of breath, while the other cluster represents more chronic causes of shortness of breath.

The two clusters discovered among patients with abdominal pain generally separate the population into a younger, healthier group and an older, sicker group.  The younger group was less likely to arrive via ambulance, and was much less likely to have previously been admitted to this hospital.  The older group was much more likely to have underlying health issues. Abdominal pain clusters are dominated by demographics and arrival mechanism.

While the abdominal pain clusters appeared more trivial in nature (old/sick, young/healthy), our workflow suggested optimized chest pain clustering would be produced with 15 neighbors used in the UMAP approximation rather than 150 neighbors.  This results in the discovered clusters having a greater reliance on local structure rather than more distant global structure.  The clusters here vary more in size, which allows for more precise phenotypes to be observed.  For instance, one of the clusters here featured patients with low risk factors for heart disease who transported themselves to the ED.  This group was less likely to need admission to the hospital.

The clusters discovered in patients presenting with back pain fit several different patient populations.  The largest cluster presents the image of patients how experience first time lower back pain, but who have good socioeconomic factors as a positive prognostic indicator.  These patients were relatively less likely to be admitted to the hospital.  The second largest cluster presents a clinical picture of patients who have medical comorbidities and non-musculoskeletal reasons for their back pain.  For instance, these patients often had leukocyturia, potentially indicating pyelonephritis (kidney infection) as the cause of their pain, or hematuria, potentially indicating nephrolithiasis (kidney stone) as the cause of their pain.  These patients were slightly more likely to be admitted to the hospital than were other patients with this chief complaint.

It is interesting to note the dissimilarity in the general shape of the training data and testing data in the patients with back pain (Figure \ref{fig:back}).  Although the training data exhibited four clusters of relatively similar density, the test data shows unbalanced clusters with much more variable density.  This is an aspect of this model where the visual nature of the embeddings can be leveraged.  While in the training data, the model easily differentiates between the blue and orange clusters (Figure \ref{fig:back_train}), the test data shows that there are a number of patients that are very similar to each of these clusters, and that these new patients are embedded with a density unlike the training densities.

The clusters discovered in patients presenting after a fall were the clusters with the highest mean pairwise ARI of any clusters within the dataset.  As shown in Figure \ref{fig:fall_all}, the resulting embeddings are very similar on subsequent training folds. These patients were split nearly in half, but split in almost the exact same way each time.  In these two clusters, the cluster with the higher admission rate is also the cluster with increased comorbidities.    These patients would likely be at increased risk of fall due to their increased age and medical comorbidities.  Similarly, falls suffered could have the potential to cause greater harm to the patient.

In sum, across the five chief complaints studied, we observed a range of phenotypes and underlying demographic and physiologic drivers. In cases of more trivial two-group separations, as observed with abdominal pain, shortness of breath, and falls, our approach reveals elements of patient comorbidities and presentation acuity. Of note, the ARIs for these complaints differ significantly, with falls (0.741) being the most stably captured and abdominal pain (0.353), the least stable. We hypothesize that this metric may be capturing the underlying heterogeneity inherent to the complaint. For example, falls may be mechanical in nature (e.g., slipping) or result from a cardiovascular cause (e.g., syncope, arrythmia) or change in mental status (e.g., transient ischemic attack). The etiologies of abdominal pain are myriad, and more difficult to bin \cite{acute_abd} - a minimal differential diagnosis of pain etiology includes a half-dozen organ systems. For this reason, abdominal pain is typically clinically assessed by abdominal quadrant. For complaints of chest and back pain, we were able to optimize for larger number of clusters. Further research is required to reveal interesting elements evolving from the pairwise comparisons of these phenotypes.

\subsection{Limitations and Future Work}
In most clinical data clusters, the visualization of the training data and the testing data appear very different.  For instance, in Figure \ref{fig:back_train}, there are well-demarcated clusters with similar appearing sizes and densities.  However, in Figure \ref{fig:back_test}, the clusters are of different sizes and appear grossly dissimilar to the clusters in Figure \ref{fig:back_train}.  The cluster on the left has nearly disappeared, while the bulk of the top and bottom clusters have become much closer to each other than in the training data.  Future work will continue to explore model robustness across variations including the use of alternative distance metrics.

Additionally, other potential clusterings of patients are likely present within these datasets.  For instance, in Figure \ref{fig:sob_ari}, it appears that another clustering might be present with 3 clusters or with 7 clusters.  Future work should evaluate these different potential clusterings to see if additional clinical characteristics can be achieved from these clusters, and should evaluate if greater number of clusters with slightly lower stability can be clinically useful.

We attempted to apply this analytic pipeline to the full clinical dataset, without separating into populations by chief complaint. We  observed that it was unable to consistently find any given clusters and hypothesize that this is likely due to the large number of potential clusters to which any given patient could belong.  Patients with the same core pathology could present with different chief complaints.  For example, a myocardial infarction (heart attack) could present as either chest pain, or as syncope.  A more generally applicable model should incorporate a method to include multiple possible chief complaints so that similar pathologies are not treated as separate entities.

An interesting extension to this work would be to utilize the clusters here to assess patients for risk of adverse events.  The clustering of patients presenting after falls exhibited characteristics that suggest that patients were grouped by their risk of falls.  The patients at lower overall risk were more likely to be discharged, while the patients at higher overall risk were more likely to be admitted.  This suggests that this technique could be applied to isolate risk factors for adverse outcomes.  This information could then help identify similar patients based on cluster membership, and allow health care providers to rapidly and effectively apply interventions to improve patient health and outcomes.

\section{Conclusion}
    \label{sec:conclusion}
This paper presents a technique for the two-dimensional visualization of complex emergency department patient data. We show that UMAP and GMMs enable robust cluster identification using a synthetic dataset and then apply these tools to a real-world clinical dataset. We explore the patient phenotypes emerging from varied patient chief complaints, revealing pertinent clinical characteristics of these populations. Among patients with abdominal pain, chest pain, shortness of breath, back pain, and falls, the populations are reliably divided into 2-6 clusters. These clusters group patients based on characteristics such as demographics and triage variables, allowing for clear clinical pictures of the type of patients involved to be seen. We anticipate future medical scenarios where deployment of this visualization pipeline will enable both rapid, real-time patient triage and retrospective cohort discovery from electronic health records.


\bibliographystyle{elsarticle-num}

\bibliography{reference}





\end{document}